\documentclass[fleqn,10pt]{wlpeerj}
\usepackage{url}

\title{Measuring Emotional Contagion in Social Media}

\author[1,2]{Emilio Ferrara}
\author[1]{Zayao Yang}
\affil[1]{School of Informatics and Computing, Indiana University, Bloomington, IN, USA}
\affil[2]{Indiana University Network Science Institute, Bloomington, IN, USA}


\begin{abstract}

Social media are used as main discussion channels by millions of individuals every day. 
The content individuals produce in daily social-media-based micro-communications, and the emotions therein expressed, may impact the emotional states of others. 
A recent experiment performed on Facebook hypothesized that emotions spread online, even in absence of non-verbal cues typical of in-person interactions, and that individuals are more likely to adopt positive or negative emotions if these are over-expressed in their social network. 
Experiments of this type, however, raise ethical concerns, as they require massive-scale content manipulation with unknown consequences for the individuals therein involved.
Here, we study the dynamics of emotional contagion using Twitter.
Rather than manipulating content, we devise a null model that discounts some confounding factors (including the effect of emotional contagion). We measure the emotional valence of content the users are exposed to before posting their own tweets.
We determine that on average a negative post follows an over-exposure to 4.34\% more negative content than baseline, while positive posts occur after an average over-exposure to 4.50\% more positive contents. 
We highlight the presence of a linear relationship between the average emotional valence of the stimuli users are exposed to, and that of the responses they produce.
We also identify two different classes of individuals: highly and scarcely susceptible to emotional contagion. 
Highly susceptible users are significantly less inclined to adopt negative emotions than the scarcely susceptible ones, but equally likely to adopt positive emotions. In general, the likelihood of adopting positive emotions is much greater than that of negative emotions.

\end{abstract}

\begin{document}

\flushbottom
\maketitle
\thispagestyle{empty}

\section*{Introduction}
The study of socio-technical systems, and their effects on our increasingly interconnected society, is playing a significant role in the emerging field of \emph{computational social science} \cite{lazer2009life,vespignani2009predicting,gilbert2009predicting,kaplan2010users,asur2010predicting,tang2012inferring,cheng2014can}.
Online social platforms like Facebook and Twitter provide millions of individuals with near-unlimited access to information and connectivity \cite{kwak2010twitter,gomez2013structure,ferrara2013traveling}.
The content produced on such platforms has proved to impact society at large: 
from social and political discussions \cite{ratkiewicz2011detecting,metaxas2012social,bond201261,conover2013digital,conover2013geospatial,varol2014evolution}, 
to emergency and disaster response \cite{sakaki2010earthquake,merchant2011integrating,lazer2014parable}, social media conversation affects the offline, physical world in tangible ways.

The central issue that inspires this work is how the content produced and consumed on social media affects individuals emotional states and behaviors. 
We are concerned in particular with the theory of emotional contagion \cite{hatfield1994emotional}.
Data from a 20-years longitudinal study suggest that emotions can be passed via social networks, and have long-term effects \cite{fowler2008dynamic}.
Various recent contributions advanced the hypothesis that emotions may be passed also via online interactions  \cite{harris2007investigation,mei2007topic,golder2011diurnal,de2012not,garcia2012positive,coviello2014detecting,coviello2014words}.
A recent study performed by Facebook suggests that emotional contagion occurs online even in absence of non-verbal cues typical of in-person interactions \cite{kramer2014experimental}.
The authors of such study performed a controlled experiment selecting a  sample of users and, by manipulating the content on their time-lines, exposed some to increased levels of positive or negative emotions, as conveyed by the posts produced by their contacts.
This experiment revealed a small but significant correlation between the number of emotionally positive/negative words in the user post and that of the stream they have been exposed to. 

The possibility to manipulate the information that users see is clearly well suited to address questions about the existence and magnitude of emotional contagion, but raises ethical concerns \cite{tufekci2014engineering,fiske2014protecting,acquisti2014economics}: the consequences of massive-scale content manipulations are unknown, and might include long-term effects on the mental and physical well-being of individuals.

In this study, we use Twitter as case study, and explore the hypothesis of emotional contagion via the social stream. 
A reasonable expectation is that Twitter connections carry a smaller emotional contagion power than Facebook ones: users generally adopt these platforms for different purposes ---Twitter for information sharing \cite{kwak2010twitter}, and Facebook to keep in touch with family and friends (or other social internetworking activities) \cite{ferrara2012large,demeo2014facebook,backstrom2014romantic}.
Yet, a recent neuroscience study found that ``\emph{reading a Twitter timeline generates 64 percent more activity in the parts of the brain known to be active in emotion than normal Web use; tweeting and retweeting boosts that to 75 percent more than a run-the-mill website}.''\footnote{This is your brain on Twitter: \url{https://medium.com/backchannel/this-is-your-brain-on-twitter-cac0725cea2b}}
In our approach we observe the Twitter stream without performing content manipulation or re-engineering of any type (no information filtering, prioritization, ranking, etc.). 
We rather devise a clever null model that discounts for emotional contagion and other correlational biases, and a method to reconstruct the stimuli (in terms of contents and their emotions) users were exposed to before posting their tweets.
This allows us to delve into the theory of emotional contagion studying single individuals and their responses to different emotions: our analysis suggests a significant presence of emotional contagion. 
We show that negative posts on average follow a 4.34\% over-exposure to negative contents prior to their production, while positive tweets occur on average after a 4.50\% over-exposure to positive contents.
We infer a linear relationship between the emotional valence of the stimuli and the response for a sample of users whose activity, and the activity of all their followees, has been monitored for an entire week during September 2014.
Our experiments highlight that different extents of emotional contagion occur: in particular, we identify two classes of individuals, namely those highly or scarcely susceptible to emotional contagion.
These two classes respond differently to different stimuli: 
highly susceptible individuals are less inclined to adopt negative emotions but equally likely to adopt positive emotions than the scarcely susceptible ones. Also, the adoption rate of positive emotions is in general greater than that of negative emotions.

Our work furthers the understanding of emotional contagion via online interactions, empowering future research with quantitative evidence of such an effect, yet at the same time avoiding the inconveniences and ethically-problematic consequences of previous experimental work carried out on other social platforms.

% You may title this section "Methods" or "Analysis". 
\section*{Materials and Methods}

%%%%%%%%%%%%%%%%%%%%%%%%%%%%%%%%
\subsection*{Sentiment Analysis}
The analysis of the emotional valence of content can be leveraged  to produce reliable forecasts in a variety of different circumstances. \cite{pang2008opinion,bollen2011twitter,bollen2011modeling,le2015predictability}.
There exists a variety of sentiment analysis algorithms able to capture positive and negative sentiment, some specifically designed for short, informal texts \cite{akkaya2009subjectivity,paltoglou2010study,hutto2014vader}.
In this work, we use SentiStrength \cite{thelwall2010sentiment,thelwall2011sentiment,stieglitz2013emotions} to annotate the tweets with positive and negative sentiment scores. Compared with other tools, SentiStrength provides several advantages: it is designed for short informal texts with abbreviations and slang (features commonly observed in Twitter), and it employs linguistic rules for negations, amplifications, booster words, emoticons, spelling corrections, particularly well suited to process social media data. SentiStrength was proven able to capture positive emotions with 60.6\% accuracy and negative emotions with 72.8\% accuracy on MySpace \cite{thelwall2010sentiment,thelwall2011sentiment,stieglitz2013emotions}.

SentiStrength assigns to each tweet $t$ a positive $S^+(t)$ and negative $S^-(t)$ sentiment score. Both scores are on a scale ranging between 1 (neutral) and 5 (strongly positive or negative). 
To capture in one single measure the sentiment expressed by each tweet, we define the \emph{polarity score} $S(t)$ as the difference between positive and negative sentiment scores assigned to tweet $t$:

\begin{equation}
	S(t) = S^+(t) - S^-(t).
	\label{eq:polarity}
\end{equation}

The polarity score $S$ ranges between -4 (extremely negative: $S^+(t)=1$ and $S^-(t)=5$) to +4 (extremely positive: $S^+(t)=5$ and $S^-(t)=1$).
When positive and negative sentiment scores for tweet $t$ are the same ($S^+(t)=S^-(t)$), we say that the polarity  of tweet $t$ is neutral ($S(t)=0$).

\begin{figure}[!t] \centering
	\includegraphics[width=\columnwidth]{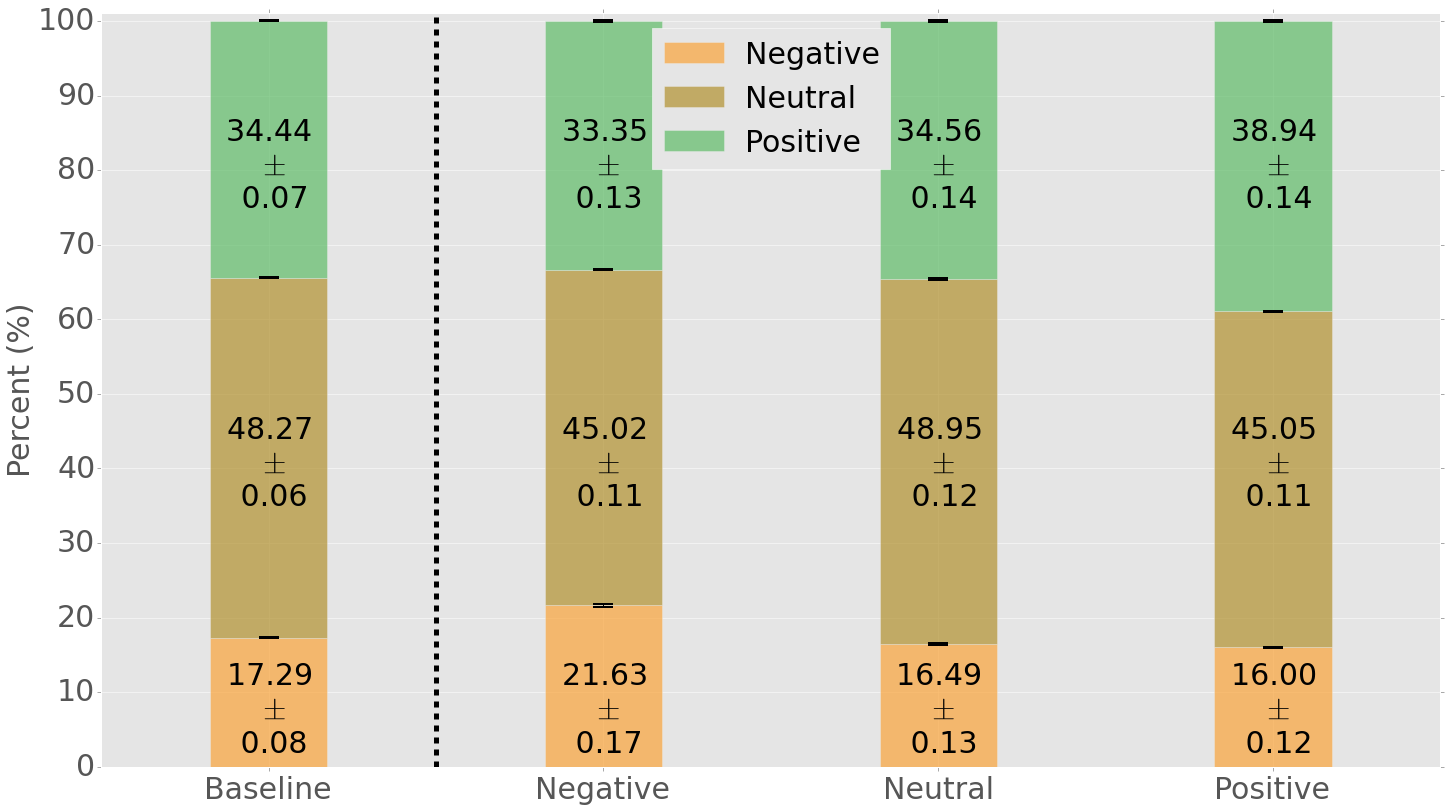} 
	\caption{{\bf Average proportions of positive, neutral, and negative emotions prior to each observed tweet.} The Baseline model (left) discounts for the effect of emotional contagion by means of a reshuffling strategy. The three bars (Negative, Neutral, and Positive) respectively show the average proportions of emotions prior to posting a negative, neutral, or positive tweet. For each negative tweet posted, on average its author was previously exposed to about 4.34\% more negative tweets than expected by the Baseline model. For each positive tweet posted, on average its author was previously exposed to about 4.50\% more positive content. Note how the distribution of emotions before posting a neutral tweet almost perfectly matches that of the Baseline model. The numbers inside the columns represent the exact proportions $\pm$ the standard errors. Error bars represent standard errors. }
	\label{fig:null_model}
\end{figure}

%%%%%%%%%%%%%%%%%%
\subsection*{Data}
Our goal is to establish a relation between the sentiment of a tweet and that of the tweets that its author may have seen in a short time period preceding its posting. To achieve that, 
we first collected a set $U$ consisting of a random sample of 3800 users who posted at least one tweet in English (among those provided by the Twitter \emph{gardenhose}) in the last week of September 2014. Via the appropriate Twitter API we also collected the set $F$ of followees of all users in $U$.

For each tweet $t$ produced by an user $u$ in $U$ in said last week of September 2014, we constructed $h_t$, the set of tweets produced by any of $u$'s followees in a time span of one hour preceding the posting of $t$.

For the purpose of our analysis we considered only tweets $t$ such that $| h_t | \geq 20$.
Also, we considered only tweets  (\emph{i}) in English, and (\emph{ii}) that do not contain URLs or media content (photos, videos, etc.). Finally, each tweet, both from the target set of users and their followees were annotated by their sentiment score as discussed above.
It is worth to briefly justify some of the choice we made. The filter English $+$ non--media was applied to be able to unambiguously attribute a sentiment score to the tweets. The choice of limiting ourselves to sampling from the last week of September was dictated by the technical limitations of the Twitter API to recover the 100\% of tweets posted by any given user only to one week prior to the query time. 
This precaution allows us to discount for possible sampling issues, so to reconstruct the full exposures to contents prior to any posting from this established set of users. Dealing with the 100\% of the content excludes possible sampling biases common to many social media studies \cite{morstatter2013sample}.
The choice to focus on tweets for which the user was exposed to at least 20 tweets within 1 hour from their posting allows us to obtain a significant description of the stimuli the users were exposed to.

We finally separated all tweets in three classes of emotions: negative (polarity score $S\leq-1$), neutral ($S=0$), and positive ($S\geq1$). 
Focusing on the classes of emotions rather than the intensity of emotions will facilitate our analysis and also discount for possible inaccuracies of the sentiment analysis procedure: several previous studies showed that it is much easier to capture the overall emotion of a short piece of text, rather than emotion intensities  \cite{akkaya2009subjectivity,paltoglou2010study,hutto2014vader,thelwall2010sentiment,thelwall2011sentiment,stieglitz2013emotions}.

% Results and Discussion can be combined.
\section*{Results}

%%%%%%%%%%%%%%%%%%%%%%%%%%%%%%%%%%%%%%%%%%%%%
\subsection*{Effect of emotional contagion}

We here want to test the hypothesis that emotional contagion occurs among social media users, as suggested by recent works on various social platforms \cite{fowler2008dynamic,kramer2014experimental}.
The idea is that emotions can be passed via online interactions even in absence of non-verbal cues typical of in-person interactions, which are deemed by traditional psychology to be an essential ingredient for emotional contagion \cite{hatfield1994emotional}.
To test this hypothesis, we need to reconstruct the emotions conveyed by the tweets each user was exposed to before posting their own tweets: this will allow us to determine whether the stimuli are correlated with the responses, namely the emotions subsequently expressed by the user.

Our study is purely observational, as we don't perform any type of controlled experiment differently from other works \cite{kramer2014experimental}. We aim to show that the average sentiment of tweets preceding a positive, negative or neutral tweet are significantly different, and determine the effect size which, even if small, at scale would have important implications.

To do so, we adopt the following reshuffling strategy aimed at determining the baseline distributions of positive, neutral, and negative contents independently of emotional contagion: for each user $u$ in the set of 3,800  users, and for each tweet $t_u$ produced by $u$, we have the history $\ell(t_u)$ of all tweets preceding $t_u$ in the 1 hour period prior to $t_u$'s publication, and we record how many such tweets $s_{\ell(t_u)}=|\ell(t_u)|$ user $u$ was exposed to.
We then put all these tweets $\ell(t_u)$, that represent the stimuli prior to the users' activities, for all tweets, for all users, in one single bucket.

To create our reshuffled null model that discounts for the effect of emotional contagion, we therefore sample with replacement\footnote{The results for sampling without replacement are substantially identical.} from bucket $B$, for each tweet $t_u$ of each user $u$, a number of tweets equal to the size $s_{\ell(t_u)}$. 
At the end of the procedure, we obtain a baseline distribution of positive, neutral, and negative sentiment prior to the publication of any tweet, which discounts for the effect of exposure and the possibility of emotional contagion.
The baseline distribution of sentiment in the null model is displayed in Fig. \ref{fig:null_model}: the proportion of positive, neutral, and negative sentiment after the exposure reshuffling is equal to, respectively, $34.44\%$ ($\pm 0.07$), $48.27\%$ ($\pm 0.06$), and $17,29\%$ ($\pm 0.08$).

\begin{figure}[!t] \centering
	\includegraphics[width=\columnwidth]{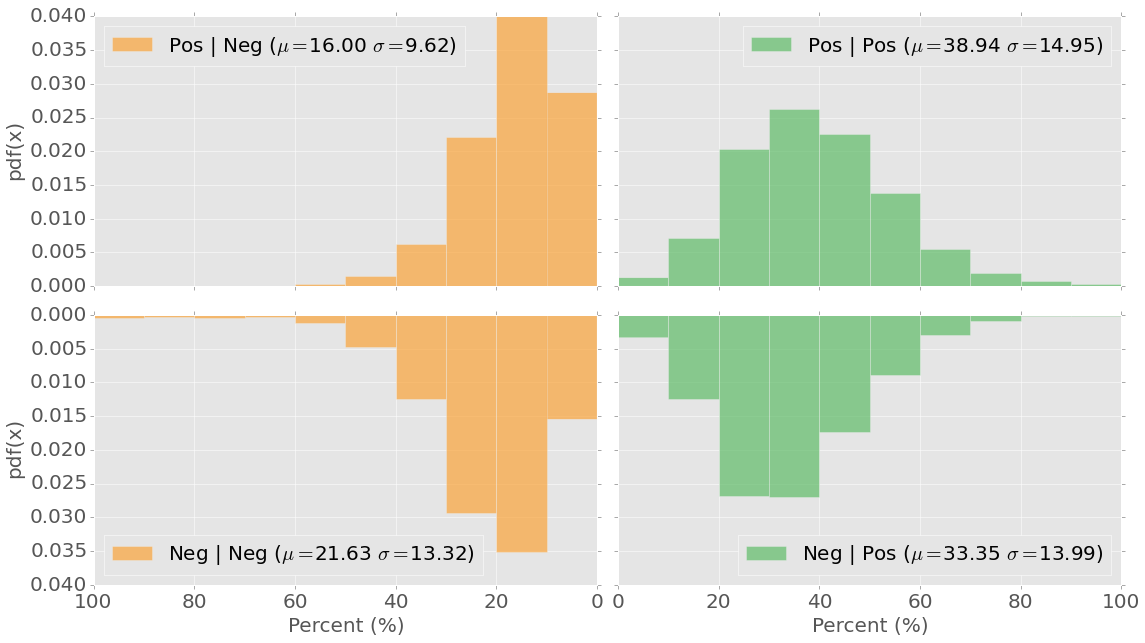} 
	\caption{{\bf Distributions of positive and negative stimuli before positive and negative responses.} The four quadrants show the probability distributions of a negative response prior to a negative (bottom left) or positive (bottom right) stimulus, or a positive response prior to a negative (top left) or positive (top right) stimulus.}
	\label{fig:stimuli_responses}
\end{figure}

To verify the hypothesis of emotional contagion, we divide all tweets $t_u$ posted by each user $u$, in three categories (positive, neutral, and negative) according to their sentiment. 
For each category, then, we generate the distribution of fraction of positive, neutral, and negative sentiments observed in the stimuli, the tweets produced by $u$'s followees prior to the posting of each $t_u$.
The results, displayed in Fig. \ref{fig:null_model}, are interpreted as follows: the three stacked-columns identify the distributions of sentiment prior to posting (from left to right) a negative, neutral, or positive tweet. 
For example, a user in our set prior to posting a negative tweet is exposed, on average, to $21.63\%$ ($\pm 0.17$) negative tweets, $45.02\%$ ($\pm 0.11$) neutral, and $33.35\%$ ($\pm 0.13$) positive ones. 
This signifies an over-exposure to $4.34\%$ more negative tweets, at the expenses of $1.09\%$ less positive ones, if compared with our null model of Fig. \ref{fig:null_model}. 
Similarly, prior to posting a positive tweet, a user in our dataset is exposed, on average, to $16.00\%$ ($\pm 0.12$) negative tweets, $45.05\%$ ($\pm 0.11$) neutral, and $38.94\%$ ($\pm 0.14$) positive ones.
This amounts for an over-exposure of $4.50\%$ more positive tweets, at the expenses of $1.29\%$ less negative ones, if compared with the null model.
Notably, the distribution of the sentiment of tweets before the posting of a neutral one matches almost perfectly the distribution of the null model in Fig. \ref{fig:null_model}, suggesting that no emotional contagion occurs in the case of neutral tweets.
To prove the statistical significance of these differences, we run a Mann–Whitney U test between the observed distributions in presence of emotional contagion, and the expected baseline of the null model. 
Both $p$ values for negative and positive emotional contagion tests are $p<10^{-6}$  while no significant difference occurs for the neutral case; the strength of the statistical significance is further illustrated by the narrow error bars in Fig. \ref{fig:null_model}.
The distributions of the positive and negative stimuli, respectively, before positive and negative responses, are also reported in Fig. \ref{fig:stimuli_responses}.

\begin{figure}[!t] \centering
	\includegraphics[width=\columnwidth]{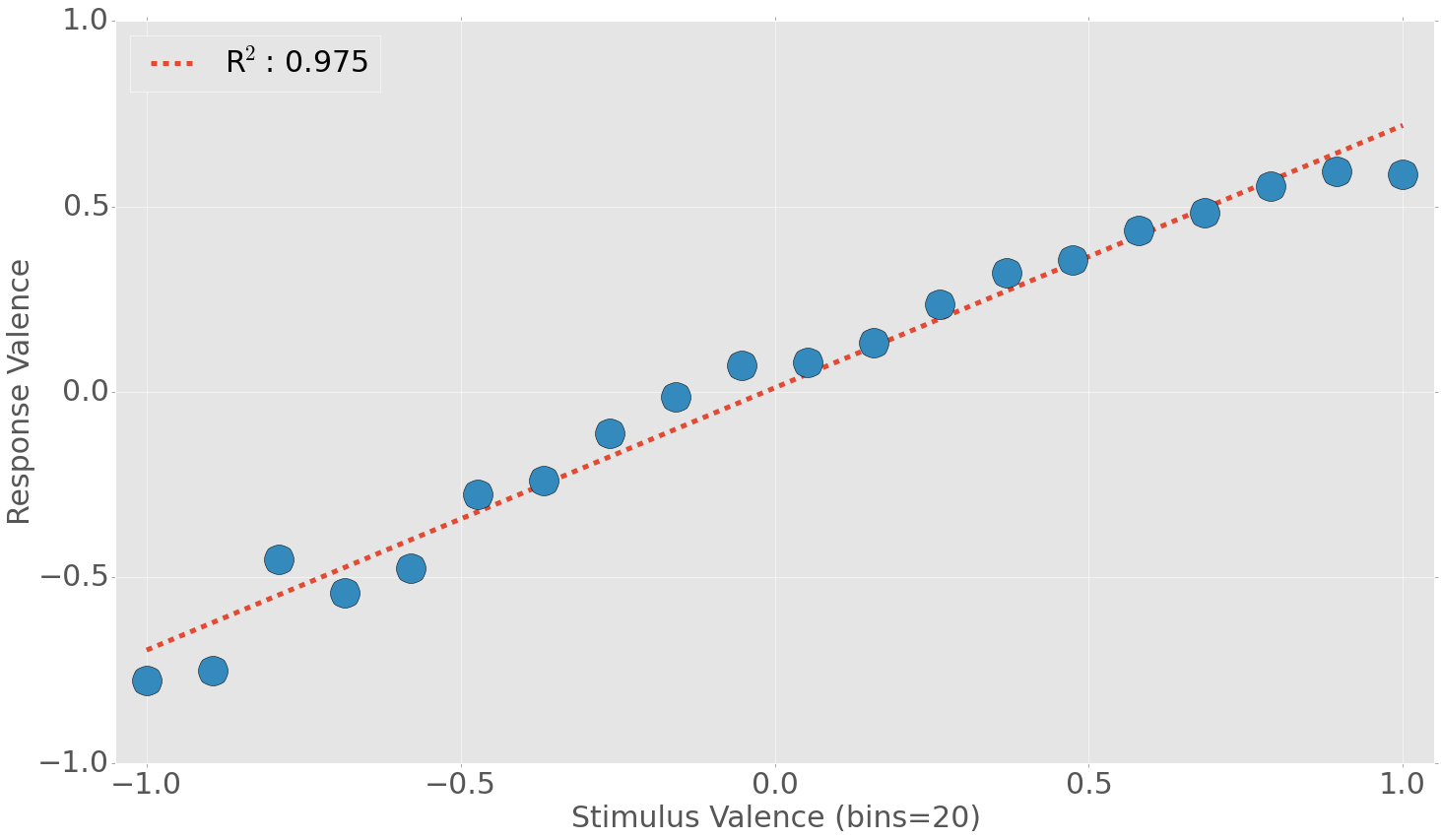} 
	\caption{{\bf Relationship between stimulus and response valence in Twitter.} The emerging linear relationship ($R^2=0.975$) suggests that there is a strong correlation between stimuli and responses in terms of valence (difference between positive and negative sentiments in the set of tweets).}
	\label{fig:contagion}
\end{figure}

These results suggest the presence of emotional contagion for both negative and positive sentiment, and seem to show that no emotional contagion occurs prior to posting neutral contents. 
To further validate this hypothesis, and in particular to focus only on positive and negative contagion, we here propose another measure, that we call \emph{valence}, that can be computed on any set (bucket) of tweets for which the sentiment is computed. Given a bucket of tweets $b$, its valence $V(b)$ is given by the following formula:

\begin{equation}
V(b) =  2 \cdot \frac{p_b}{n_b + p_b}-1 
\label{eq:valence}
\end{equation}

where $p_b$ and $n_b$ represent, respectively, the fraction of positive and negative tweets in bucket $b$. This measure ranges between -1 and +1: the lower the score, the larger the disproportion toward negative emotion, and vice-versa.

Since for each tweet $t_u$ produced by each user $u$ we already obtained the history $\ell(t_u)$ of all tweets preceding $t_u$ in the 1 hour period prior to $t_u$'s publication, we can compute the valence scores $V(\ell(t_u))$ for all histories. This allows us to represent the difference in intensity between positive and negative stimuli each user $u$ was exposed to prior to posting each tweet $t_u$. 
Therefore, we calculate the valence scores $V(\ell(t_u))$ for all tweets $t_u$ in our dataset. 
This generates a distribution of values between -1 and +1, each value representing the valence of the stimulus of the associated tweet. We then bin these \emph{stimuli valence} values, in 20 bins of length 0.05 (see the $x$-axis of Fig. \ref{fig:contagion}.) Each bin $x_b$ contains, again, a set of tweets (the responses) for which we already calculated the sentiment (positive, negative, or neutral). We can calculate also the valence of each $x_b$. Such values will represent the \emph{response valence} for a given value (bin) of stimulus valence.

The results, illustrated in Fig. \ref{fig:contagion}, show a very strong linear relationship ($R^2=0.975$) between the valence of the stimulus and the valence of the response. For example, a very strong negative stimulus with valence -1 generates a response valence of about -0.8. Similarly, a very strong positive stimulus of valence +1 will trigger a response of valence around +0.6.
These results suggest a common mechanism of contagion in both negative and positive contents: in general, a strongly negative stimulus is followed by negative responses, while a strongly positive stimulus generates positive responses. Neutral stimuli also trigger neutral responses.

\subsection*{Extent of emotional contagion and individuals' susceptibility}

\begin{figure}[!t] \centering
	\includegraphics[width=\columnwidth]{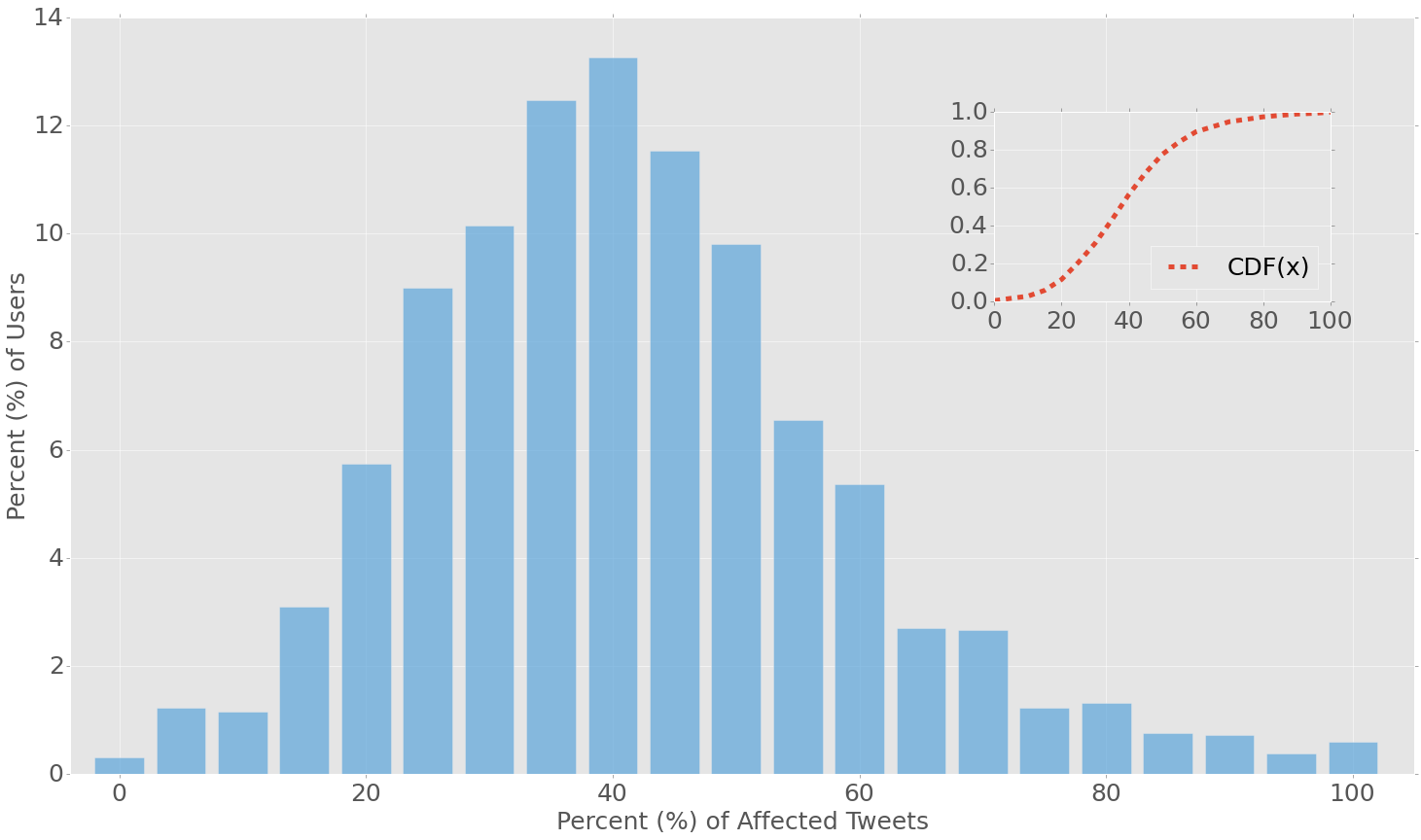} 
	\caption{{\bf Measurement of emotional contagion on users' content posted on Twitter.} The main plot shows the number of users as function of the fraction of their tweets affected by emotional contagion. The inset shows the cumulative distribution. About 80\% of the users have up to 50\% of their tweets affected by emotional contagion, while the remainder 20\% of users exhibits effects of emotional contagion on more than 50\% of the posts they produce.}
	\label{fig:contagion_extent}
\end{figure}

Using data collected for the previous experiment, we can also explore if different users have different susceptibility to emotional contagion, for example by measuring how many of their tweets reflect the over-represented stimulus prior to the postings.
We now focus our attention on the tweets posted by each of the 3,800 users in our dataset and all tweets produced by their followees.

To determine whether user $u$ was susceptible to emotional contagion prior to posting any of her/his tweets, for each tweet $t_u$ posted by $u$ we calculate the proportions of positive $p^+$, neutral $p^\circ$, and negative $p^-$ polarities computed from the distribution of all tweets produced by $u$'s followees in the 1 hour prior to $t_u$'s posting time. 
This triplet $O=\{p^+, p^\circ, p^- \}$ has three entries that indicate the proportion of each of the three sentiment states $\{+, \circ, - \}$.
These tweets are considered as the stimulus to which user $u$ was exposed prior to posting tweet $t_u$.  

The following baseline proportions are derived by the previous experiment (Fig. \ref{fig:contagion}): 

$$B^- = \{ 21.63, 45.02, 33.35 \}$$
$$B^\circ = \{ 16.49, 48.95, 34.56  \}$$
$$B^+ = \{ 16.00, 45.05, 38.94 \}.$$

We therefore determine the smallest Euclidean distance $\bar{L_2}(O, B^s)$ among the distances between the observed distribution $O$, and any of the three baseline sentiment proportions $B^-$, $B^\circ$, and $B^+$.
This to determine the nature of the stimulus to which $u$ is exposed prior to posting (over-exposure to negative, neutral, or positive):

\begin{equation}
	\bar{L_2}(O, B^s) = \min_{s \in \{+, \circ, -\}} \; \left\{ \sqrt{ \sum_{i \in \{+, \circ, - \}}{(O_i - B_i^s)^2} } \right\} \;
	\label{eq:distance}
\end{equation}
\\

\begin{figure}[!t] \centering
	\includegraphics[width=\columnwidth]{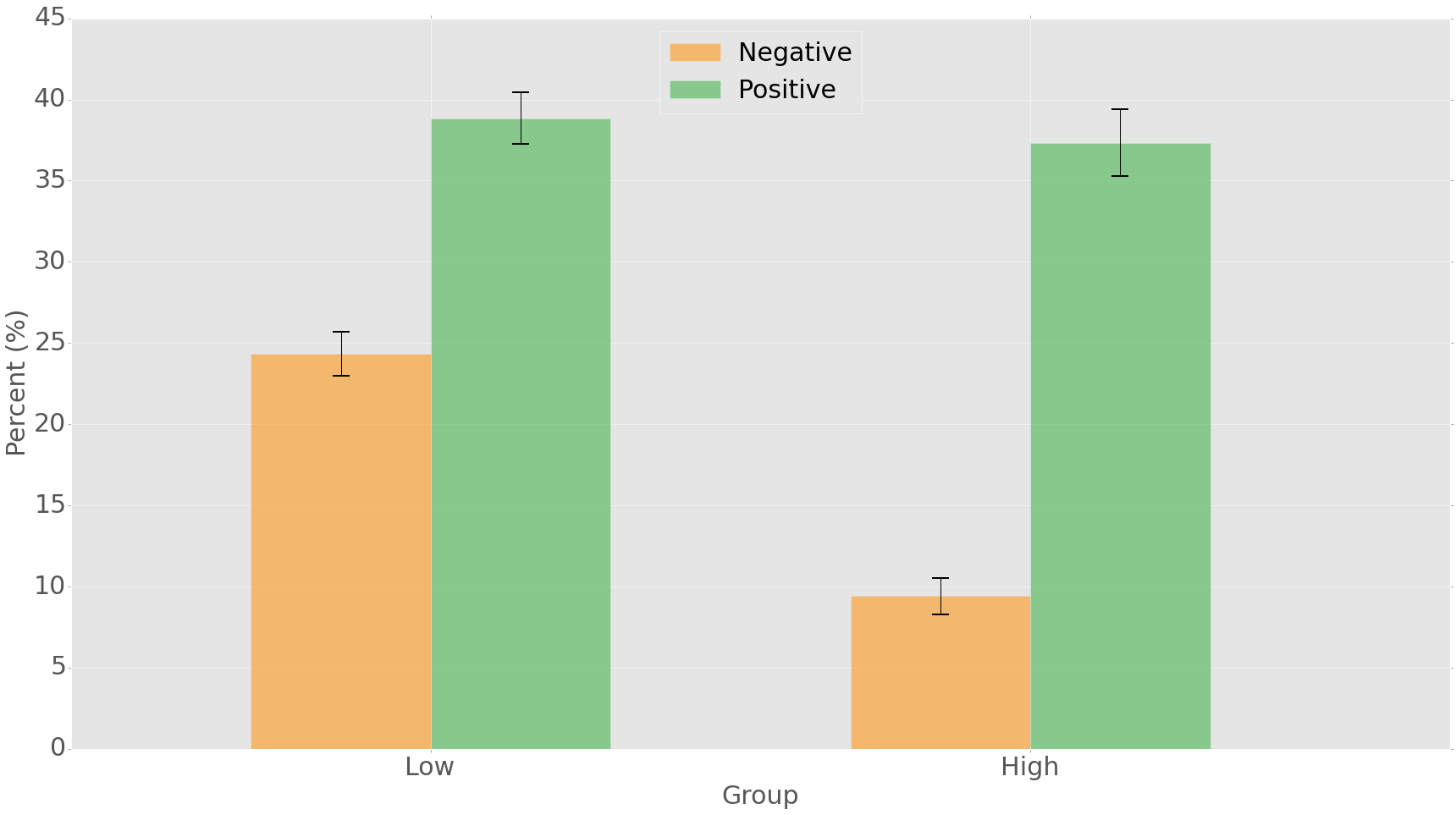} 
	\caption{{\bf Different extent of emotional contagion on the two groups of scarcely and highly susceptible users.} Highly susceptible users are significantly less inclined to adopt negative emotions than the scarcely susceptible ones, but equally likely to adopt positive emotions. In general, the likelihood of adopting positive emotions is much greater than that of negative emotions. 	}
	\label{fig:contagion_types}
\end{figure}

If the smallest distance is, say, $\bar{L_2}(O, B^-)$ it means that, in presence of emotional contagion, $u$ would be expected to post a negative tweet given the over-exposure to negative content. 
Similarly, if the smallest distance is $\bar{L_2}(O, B^+)$, then $u$ is expected to post a positive tweet, in case s/he being affected by emotional contagion.
If $u$ were to tweet according to the stimuli (s)he is exposed to, then we consider $t_u$ to be outcome of susceptibility to emotional contagion; vice-versa, $t_u$ is counted as instance of $u$ being insusceptible to emotional contagion given the stimuli.

We perform this analysis for all tweets of all users, and characterize each user $u$ with a fraction summarizing the proportion of tweets affected by emotional contagion. 
Fig. \ref{fig:contagion_extent} shows the distribution of this measure for all users (the inset of Fig. \ref{fig:contagion_extent} illustrates the cumulative distribution): it is evident that about 80\% of the users have up to 50\% of their tweets affected by emotional contagion, while the remainder 20\% exhibits very high susceptibility and demonstrate that more than 50\% of the content they post suggests the presence of emotional contagion.

We further divide the users in two categories, highly and scarcely susceptible to emotional contagion, by selecting the top and bottom 15\% of the distribution, respectively.
For these two classes independently we compute the fraction of susceptible tweets that are positively or negatively affected by emotional contagion, we average these fractions across users, and we plot the results in Fig. \ref{fig:contagion_types}.
We can note that two very different emotional contagion dynamics exist: the group of users who are more susceptible to emotional contagion, are significantly more inclined to adopt positive emotions rather than negative. 
The vice-versa happens for users scarcely susceptible to emotional contagion: they adopt much more frequently negative emotions in the uncommon occurrences when they are susceptible to emotional contagion.
However, the probability of a contagion of positive emotions is much greater than the negative case in both susceptibility classes: the low- and high-susceptibility groups are, respectively, 1.6 times and 3.96 times more likely to adopt positive emotions, with respect to negative ones.

\section*{Discussion}
In this study we performed an extensive observational analysis of the patterns of emotional contagion on Twitter. Differently from a  study carried out on Facebook \cite{kramer2014experimental}, where controlled experiments were performed to manipulate the exposure to arbitrary emotions, in this study we observe and measure emotional contagion without interacting with the users. The design of a clever null model, which discounts some confounding factors including contagion, allows us to highlight the effect of emotional contagion on a sample of 3,800 users whose entire history of stimuli and responses has been tracked throughout a week of activity.
Our results suggest a number of insights: we can hypothesize the presence of emotional contagion even without the hassle (and ethical concerns) of manipulating users time-lines. We observed that, on average, a negative tweet follows an over-exposure to 4.34\% more negative stimuli, whereas a positive one follows an over-exposure to 4.50\% more positive tweets. A strong linear relation emerges between the valence of the stimuli and that of the responses, suggesting that a common mechanism of contagion exists regulating both negative and positive emotions.
Finally, by dividing the users in two categories (highly and scarcely susceptible), we observed how, in general, positive emotions are more prone to contagion, and that highly-susceptible users are significantly more inclined to adopt positive emotions.

Our study is certainly not immune of possible shortcomings: theoretical work by Shalizi and Thomas \cite{shalizi2011homophily}, for example, suggests that in observational studies like ours it is not possible to separate contagion from homophily. In a world entirely dominated by homophily, our observation would not imply an effect: users prone to produce negative contents would link only to others with same emotional alignment (and vice-versa for positive-inclined ones). However, in  a real world where it makes sense to assume a mixture of contagion and homophily dynamics, our experiments suggest the presence and the extent of emotional contagion, while further work will be needed to understand the effect of homophily and how that intertwines with emotional contagion.

\bibliographystyle{abbrv}
\bibliography{references}

\end{document}